\documentclass[10pt]{article}
\usepackage{authblk}
\usepackage[T1]{fontenc}
\usepackage{latexsym}
\usepackage{physics}
\usepackage{amsmath}
\usepackage{amssymb}
\usepackage[english]{babel}
\usepackage{array}
\usepackage{verbatim} 
\usepackage{graphicx}
\usepackage{footmisc}
\makeatletter
\newcommand*\bigcdot{\mathpalette\bigcdot@{.5}}
\newcommand*\bigcdot@[2]{\mathbin{\vcenter{\hbox{\scalebox{#2}{$\m@th#1\bullet$}}}}}
\makeatother

\newcommand{\di}{\displaystyle}
\newcommand{\qul}{\textquotedblleft}
\newcommand{\qur}{\textquotedblright}

\begin{document}
	
	\renewcommand{\refname}{Comments with References}
	
	\title{\textbf{On the Gravitational Effects of Light}}

	\author{Alessio Rocci\thanks{alessio.rocci@vub.be}}
	\affil{Department of Physics Theoretische Natuurkunde (VUB) and the International Solvay Institutes; Applied Physics research group (APHY)}


\maketitle

	\abstract{\textbf{Translation of and Commentary on Léon Rosenfeld's “\"Uber die Gravitationwirkungen des Lichtes”, \textit{Zeitschrift fur Physik} 65: 589–599 (1930). Originally published in German. Submitted for publication on September 26, 1930. See \cite{preface} in the \textit{Comments with References} section before reading this English translation.}\\ The gravitational field generated by an electromagnetic field is calculated using of laws of quantum mechanics and it is shown that the resulting gravitational energy turns out to be infinitely large, raising a new difficulty for the Heisenberg-Pauli quantum theory of wave fields. In addition, the transition processes in first-order approximation involving light and gravitational quanta \cite{graviton} are briefly discussed.}


	As is well known\footnote{W. Heisenberg, and W. Pauli, Zeitschrift f\"ur Physik \textbf{56}, 1, 1929; R. Oppenheimer, Physical Review \textbf{35}, 461, 1930; I. Waller, Zeitschrift f\"ur Physik \textbf{62}, 673, 1930.} \cite{footnote}, the occurrence of an infinitely large self-energy of the electron causes severe difficulties for quantum electrodynamics \cite{self-energy-electron}.
	Heisenberg posed the question whether, regardless of any interaction with matter, analogous behavior occurs also in the case of the gravitational effects of light \cite{Heisenberg}. The answer cannot be obtained simply by comparing the latter with the case of the electron, since here the retardation effects cannot be neglected \cite{retardation}. The present work examines this question \cite{quantization}.

	\section{The gravitational field generated by an electromagnetic field in first order approximation\protect\footnote{A. Einstein Berliner Berichte 1918, p. 154, or W. Pauli, Relativit\"atstheorie, Berlin 1921, n. 60, p. 736.}}
	Let $ \varkappa = \frac{8\pi G}{c^4} $ be Einstein's gravitational constant \cite{Newton-Constant} ($ G =$ Newton's constant) and let us assume that the gravitational field components deviate so little from their Minkowski value \cite{weak-field} that we can expand them in powers of $ \varepsilon=\sqrt{\varkappa} $ and only need to take into account the terms linear in $ \varepsilon $. In Cartesian coordinates \cite{old-time-coordinate} $ x^1,\, x^2,\, x^3,\, x^4=ict $, we can write:
	\begin{equation}
		g_{ik} = \delta_{ik} + \varepsilon\gamma_{ik}\; . 
	\end{equation}
	If we set
	\begin{equation}
		\gamma = \sum_{i}\gamma_{ii}   
	\end{equation}
	and
	\begin{equation}
		\gamma'_{ik} = \gamma_{ik}-\frac{1}{2}\delta_{ik}\gamma \, ,
	\end{equation}
	it follows that
	\begin{eqnarray}
		\gamma' &=& -\gamma \\
		\gamma_{ik} &=&\gamma'_{ik}-\frac{1}{2}\delta_{ik}\gamma' \, .
	\end{eqnarray}
	We can then determine the coordinate system through the requirement\footnote{The conventional rule of omitting the summation sign has been adopted wherever it does not affect clarity.} \cite{derivative-notation}
	\begin{equation}\label{gauge}
		\partial_k\gamma'_{ik}=0\, .
	\end{equation}
	The electromagnetic field $ F^{ik} $ with Maxwell's energy-momentum tensor \cite{F-squared}
	\begin{equation}\label{em-stress-energy}
		\left.
		\begin{aligned}
			S_{ik} &= F_{ir}F_{kr}-\frac{1}{4}\delta_{ik}F^2 \\
			F^2 &= F_{rs}F_{rs} 
		\end{aligned}
		\right\} 
	\end{equation}
	generated by the gravitational field is given by the following equation \cite{dalembertian}
	\begin{equation}\label{eq8}
		\square \gamma'_{ik} = -2\varepsilon  S_{ik}  \, .
	\end{equation}
	Incidentally, from $ \di \Sigma_i S_{ii}=0 $ it follows that
	\begin{equation}
		\square \gamma' = 0
	\end{equation}
	and hence we can also write
	\begin{equation}\label{eq10}
		\square \gamma_{ik} = -2\varepsilon  S_{ik} 
	\end{equation}
	instead of eq. (\ref{eq8}). In addition, we should consider also Maxwell's equations modified by the gravitational terms that do not need to be written down explicitly.
	The corresponding Lagrange function is \cite{Lagrangian}:
	\begin{equation}\label{Lagrangian}
		\mathcal{L} = -\frac{1}{4}F^2-\frac{1}{8}\left( \partial_i\gamma'_{rs}\partial_i\gamma'_{rs} -\frac{1}{2}\partial_i\gamma'\partial_i\gamma'\right) +\frac{\varepsilon}{2}\gamma'_{rs}S_{rs}\, .
	\end{equation}
	From eq. (\ref{Lagrangian}) follows the expression for the Hamiltonian, which we write using $ q $ and $ q' $ variables (instead of $ q $ and $ p $) \cite{variables}, namely:
	\begin{equation}\label{Hamilt-tot}
		\mathcal{H} = \mathcal{H}_{L} + \mathcal{H}_G + \mathcal{W},
	\end{equation}
	where $ \di \mathcal{H}_{L} = \frac{1}{2}\left( \vec{E}^2 + \vec{H}^2 \right)  $ is the usual electromagnetic energy density,  $ \di \mathcal{H}_{G} $ is the pure gravitational part, namely \cite{time-derivative}
	\begin{equation}\label{hamilt-G} 
		\mathcal{H}_G=\frac{1}{8}\left(\partial_i\gamma'_{rs}\partial_i\gamma'_{rs}-\frac{1}{2}\partial_i\gamma'\partial_i\gamma' \right) -\frac{1}{4}\left( \dot{\gamma'}_{rs}\dot{\gamma'}_{rs}-\frac{1}{2}\dot{\gamma'}\dot{\gamma'} \right)  ,
	\end{equation}
	and $ \mathcal{W} $ is the interaction part\footnote{\label{mista} We fixed (for the electromagnetic potential) $ \Phi_4 = 0 $ \cite{gauge}; compare with W. Heisenberg and W. Pauli, Zeitschrift f\"ur Physik 59, 171ff., 1930.}:
	\begin{equation}\label{H-int} 
		\mathcal{W}=-\frac{\varepsilon}{2}\left( \gamma'_{rs}S_{rs}-2\gamma'_{4s}S_{4s}-2\gamma'_{is}F_{4i}F_{4s}-\frac{1}{2}\gamma'_{44}F^2 + \gamma'F_{4i}F_{4i}\right)\, . 
	\end{equation}
	
	When quantizing the $ \gamma'_{ik} $, we shall take into account the condition (\ref{gauge}) using "Fermi's method"\footref{mista}. This means that we shall impose this condition and its time derivative on a $ t= constant$ hypersurface. It should be noticed that they are not allowed to be understood as q-numbers relations. It is then easy to see, taking into account the conservation law $ \partial_k S_{ik}=0 $, that the relevant constraints propagate over time due to the field equations (\ref{eq8}) \cite{Fermi}.

	We will treat the electromagnetic field (zeroth-order) using the second method presented by Heisenberg and Pauli\footref{mista}. Since we are not considering matter, we are concerned only with transversal degrees of freedom \cite{vibration}; we introduce the plane wave decomposition of the electromagnetic field choosing periodic boundary conditions with period $ L $
	and letting $ L $ ultimately go to infinity\footnote{L. Landau and R. Peierls, Zeitschrift f\"ur Physik \textbf{62}, p. 197, 1930.}. Thus, we set \cite{electro-magnetic-field}
	\begin{eqnarray}\label{em-fields}
		\vec{E} &=& \frac{\alpha}{\sqrt{L^3}}\sum_{\vec{k},\lambda}\sqrt{\frac{k}{L}}\;\vec{e}_{\vec{k}\lambda}\left( A_{\vec{k},\lambda}\, e^{\frac{2\pi i\vec{k}\bigcdot\vec{x}}{L}}- A_{-\vec{k},-\lambda}\, e^{-\frac{2\pi i\vec{k}\bigcdot\vec{x}}{L}}\right) \nonumber\\
		\vec{H} &=& \frac{\alpha}{\sqrt{L^3}}\sum_{\vec{k},\lambda}\sqrt{\frac{k}{L}}\;\vec{h}_{\vec{k}\vec{x}}\left( A_{\vec{k},\lambda}\, e^{\frac{2\pi i\vec{k}\bigcdot\vec{x}}{L}}- A_{-\vec{k},-\lambda}\, e^{-\frac{2\pi i\vec{k}\bigcdot\vec{x}}{L}}\right)
	\end{eqnarray}
	where $ \di \alpha = \frac{1}{i}\sqrt{\frac{ch}{2}} $ is a normalization factor; the index $ \lambda = 1,2 $ labels the two perpendicular polarizations \cite{polarization} with wave propagation vector $\vec{k}$ and modulus $k=\abs{\vec{k}}$ \cite{propagation}; $\vec{e}_{\vec{k}\lambda}$ and  $\vec{h}_{\vec{k}\lambda}$ are both normalized vectors, which are perpendicular to $\vec{k}$ and to each other, and they satisfy $ \di \vec{h}_{\vec{k}\lambda} = \vec{e}_{\vec{k}\lambda}\times \vec{k}/k  $. Furthermore $ \vec{e}_{\vec{k}\lambda}=\vec{e}_{-\vec{k}\lambda} $ and $ \vec{h}_{-\vec{k}\lambda}=-\vec{h}_{\vec{k}\lambda} $. Finally, we write the amplitudes $ A $ using the \textit{number-} and \textit{phase-}variables as follows \cite{conjugated-variables}:
	\begin{equation}\label{eq16}
		A_{\vec{k},\lambda} = e^{-\frac{2\pi i }{h}\Theta_{\vec{k}\lambda}}\; \sqrt{N_{\vec{k}\lambda}}\;\; ,\qquad A_{-\vec{k},-\lambda} = \sqrt{N_{\vec{k}\lambda}}\; e^{\frac{2\pi i }{h}\Theta_{\vec{k}\lambda}}\; .
	\end{equation}
	For the sake of clarity, in the following we will denote a specific state $ (\vec{k}_r,\,\lambda_r) $ by the letter $ r $ and set accordingly $ A_{\vec{k}_r,\,\lambda_r} \equiv A_r$, $ A_{-\vec{k}_r,\,-\lambda_r} \equiv B_r$. Furthermore, we introduce the following abbreviations \cite{shortcuts}:
	\begin{equation}
		\left.
		\begin{aligned}
			I_{\pm 5}(rs) &= \frac{1}{L^3}e^{\frac{2\pi i}{L}(\vec{k}_r\pm\vec{k}_s)\bigcdot\vec{x}} \\
			I_{\pm 6}(rs) &=  \frac{1}{L^3}\frac{e^{\frac{2\pi i}{L}(\vec{k}_r\pm\vec{k}_s)\bigcdot\vec{x}}}{\abs{\vec{k}_r\pm\vec{k}_s}^2-(k_r\pm k_s)^2}\frac{L^2}{2\pi^2}\\
			&= \pm \frac{L^2}{4\pi^2}\frac{1}{k_rk_s(\text{cos}\,\theta_{rs}-1)}I_{\pm 5}(rs)  \\
			&\qquad \left(  \text{cos}\,\theta_{rs} = \frac{\vec{k}_r\bigcdot\vec{k}_s}{k_rk_s} \right) \, , \\
			I_{\pm i}(rs) &= \frac{\partial I_{\pm 6}(rs)}{\partial x^i}= \frac{2\pi i}{L}(\vec{k}_r\pm\vec{k}_s)_i I_{\pm 6}(rs)\qquad (i=1,2,3)\, , \\
			I_{\pm 4}(rs) &= -\frac{2\pi}{L}(k_r\pm k_s)I_{\pm 6}(rs) \, ,  \\
			I_{\pm 4}^*(rs) &= \frac{2\pi}{L}(k_r\pm k_s)I_{\pm 6}^*(rs) \, ,
		\end{aligned}
		\right\}
	\end{equation}
	($ x^* = $ complex conjugate of $ x $).

	The above explanations only apply if $ \vec{k}_r $ is not parallel to $ \vec{k}_s $; as we shall see, this special case drops out automatically in the following computations. Finally, we define the tensor $ \mathfrak{s}^{rs}_{ik} $ as follows \cite{scalar}:
	\begin{eqnarray}\label{eq18}
		-\mathfrak{s}^{rs}_{44} &\equiv& \mathfrak{w} = \frac{1}{2}\sum_{i} (e^r_ie^s_i + h^r_ih^s_i) \nonumber\\
		\mathfrak{s}^{rs}_{il} &=& \delta_{il} \mathfrak{w} - \frac{1}{2}(e^r_ie^s_l + h^r_ih^s_l) - \frac{1}{2}(e^s_ie^r_l + h^s_ih^r_l)   \nonumber\\
		\mathfrak{s}^{rs}_{i4} &=& \mathfrak{s}^{rs}_{4l} =  \frac{i}{2} [(\vec{e}_r\times\vec{h}_s)_l +(\vec{e}_s\times\vec{h}_r)_l]\quad (i,l = 1,2,3)\, .
	\end{eqnarray}
	Hence, by using our notation, eq. (\ref{em-fields}) and eq. (\ref{em-stress-energy}) imply that
	\begin{multline}\label{em-stress-energy-2}
		S_{ik} =  \alpha^2\sum_{rs}\sqrt{\frac{k_rk_s}{L^2}}\mathfrak{s}^{rs}_{ik}\cdot \left\lbrace A_rA_sI_{+5}(rs) +B_rB_sI^*_{+5}(rs)\right.\\
		\left. -A_rB_sI_{-5}(rs)-B_rA_sI^*_{-5}(rs)\right\rbrace \; ;
	\end{multline} 
	where we assume that only the gravitational field produced by the light field (\ref{em-stress-energy-2}) is present \cite{self-interaction}. Therefore, by considering the phases of $ A $, $ B $ and the periodic boundary conditions, the gravitational field solving eq. (\ref{eq10}) is:
	\begin{multline}\label{eq20}
		\gamma_{ik} = \varepsilon\alpha^2\sum_{rs}\sqrt{\frac{k_rk_s}{L^2}}\mathfrak{s}^{rs}_{ik}\cdot
		\left\lbrace A_rA_sI_{+6}(rs)+B_rB_sI^*_{+6}(rs)\right.\\
		\left. - A_rB_sI_{-6}(rs)-B_rA_sI^*_{-6}(rs)\right\rbrace \; .
	\end{multline}  
	This expression, which is a valid solution also when the variables are q-numbers \cite{q-numbers}, satisfies the constraints (\ref{gauge}) due to the conservation law
	\begin{equation}
		\partial_k S_{ik} = 0\; .\nonumber
	\end{equation} 
	Furthermore, it is easy to see that
	\begin{equation}\label{eq21}
		\lim_{L\rightarrow\infty} I_{-6}(rr) = 0
	\end{equation}
	because it plays the role of a (retarded) potential of constant density $ 1/L^3 $,  vanishing everywhere at the limit $ L\rightarrow\infty $.

	\section{Computing of the gravitational energy}\label{computation}
	As the gravitational field (\ref{eq20}) is first order in $ \varepsilon $, the gravitational energy $ \di \int\left( \mathcal{H}_G+\mathcal{W}\right) dV $ of eq. (\ref{hamilt-G}) and (\ref{H-int}) is the second order $ \varepsilon^2 $. In perturbation theory, the appropriate correction to the energy (i.e. to the second order) \cite{energy} $ \di \overline{\mathcal{H}}_L = \sum_r(N_r+\frac{1}{2})h\nu_r $ (where $ \nu_r=\frac{k_rc}{L} $) of a specific state of the electromagnetic field, which corresponds to considering the number $ N_r $ of light quanta labelled by $ r $, can be obtained by inserting equation (\ref{eq20}) into the expression of $ \mathcal{H}_G+\mathcal{W} $ and by calculating the contribution of the corresponding diagonal element for the relevant state \cite{tad-pole}.

	A first simplification is based on noticing that for the field (\ref{eq20}) $ \gamma=-\gamma'=0 $ and thus $ \gamma_{ik}=\gamma_{ik}' $. The remaining terms have the following forms:
	\begin{equation}\label{eq22}
		\begin{aligned}
			\varepsilon^2&\alpha^4\sum_{rsmn}\sqrt{\frac{k_rk_sk_mk_n}{L^4}}\mathfrak{s}^{rs}_{ik} \mathfrak{s}^{mn}_{ik}\cdot\\
			&\int \left\lbrace A_rA_s I_{+\tau}(rs) + B_rB_s I^*_{+\tau}(rs) - A_rB_s I_{-\tau}(rs)
			- B_rA_s I^*_{-\tau}(rs) \right\rbrace\cdot\\
			&\left\lbrace A_mA_n I_{+\rho}(mn) + B_mB_n I^*_{+\rho}(mn)
			- A_mB_n I_{\rho}(mn) - B_mA_n I^*_{-\rho}(mn)\right\rbrace  dV \\
			&(\rho,\tau = 1,2,\dots,6)
		\end{aligned}
	\end{equation}
	or similar expressions where $ \mathfrak{s}^{rs}_{ik}\mathfrak{s}^{mn}_{ik} $ is replaced by
	\begin{equation}\label{eq23}
		\mathfrak{s}^{rs}_{i4}\mathfrak{s}^{mn}_{i4}-\frac{1}{2}\mathfrak{s}^{rs}_{ik}(e_i^me_k^n+e_i^ne_k^m)\, ,
	\end{equation}
	or 
	\begin{equation}\label{eq23b}
		(\vec{h}_r\bigcdot\vec{h}_s + \vec{e}_r\bigcdot\vec{e}_s)(\vec{h}_m\bigcdot\vec{h}_n - \vec{e}_m\bigcdot\vec{e}_n)\; .\tag{23\textquotesingle}
	\end{equation}
	In the integrand of eq.(\ref{eq22}), only products with two $ A $ factors and two $ B $ factors with the same index contain diagonal terms. Utilizing eq. (\ref{eq21}), the expression (\ref{eq22}) reduces to the following form:
	\begin{multline}\label{eq24}
		\varepsilon^2\alpha^4\sum_{rs}\frac{k_rk_s}{L^2}
		\sum_{ik}(\mathfrak{s}^{rs}_{ik})^2
		\int dV\;\;  \left\lbrace   
		2A_rA_sB_rB_s I_{+\tau}(rs)I_{+\rho}^*(rs)\right.\\
		+ 2B_rB_sA_rA_s I^*_{+\tau}(rs)I_{+\rho}(rs)
		+ A_rB_sA_sB_r I_{-\tau}(rs)I_{-\rho}(sr) \\
		+ A_rB_sB_rA_s I_{-\tau}(rs)I^*_{-\rho}(rs)
		+ B_rA_sA_rB_s I^*_{-\tau}(rs)I_{-\rho}(rs)\\
		\left. + B_rA_sB_sA_r I^*_{-\tau}(rs)I^*_{-\rho}(sr) \right\rbrace \, .
	\end{multline}
	To easily evaluate the expressions for $ \sum_{ik}(\mathfrak{s}^{rs}_{ik})^2 $ and eq. (\ref{eq23}), we observe that they are covariant under rotation. By choosing therefore $ \vec{h}_s $, $ \vec{e}_s $ and $ \vec{k}_s $ as a reference system, the tensor  $ \mathfrak{s}^{rs}_{ik} $ of eq. (\ref{eq18}) reduces to
	\begin{equation}
		\left.
		\begin{matrix} 
			\di	\frac{1}{2}(e_2-h_1)\, ,     &\di -\frac{1}{2}(e_1+h_2)\, , &\di -\frac{1}{2}h_3\, ,		 & \di		\frac{i}{2}h_3\, , \\[0,5cm]
			\di	-\frac{1}{2}(e_1+h_2) \, ,   & \di-\frac{1}{2}(e_2-h_1)\, ,  &\di -\frac{1}{2}e_3\, ,		 & \di \frac{i}{2}e_3 \, ,     \\[0,5cm]
			\di	 - \frac{1}{2}h_3\, , &\di -\frac{1}{2}e_3 \, ,	& \di \frac{1}{2}(e_2+h_1) \, ,&\di	-\frac{i}{2}(e_2+h_1)\, , \\[0,5cm]
			\di	 \frac{i}{2}h_3 \, ,	     & \di\frac{i}{2}e_3\, ,		&\di -\frac{i}{2}(e_2+h_1)\, ,	&\di -\frac{1}{2}(e_2 + h_1)\, ,\nonumber
		\end{matrix} 
		\right\}
	\end{equation}
	This implies, by using the orthogonality relations, that the following expressions are related by a proportionality constant \cite{duepunti}:
	\begin{eqnarray}
		\sum_{i,\, k}(\mathfrak{s}^{rs}_{ik})^2\; &:&\; \frac{1}{2}\left( 1-\frac{k_3}{k} \right)^2 \nonumber\\
		\sum_{i}(\mathfrak{s}^{rs}_{i4})^2 - \frac{1}{2}\mathfrak{s}^{rs}_{ik}(e_i^re_k^s+e_i^se_k^r) \; &:&\; \frac{1}{4}\left( 1-\frac{k_3}{k} \right)^2 \nonumber\, .
	\end{eqnarray}
	On restoring the original coordinate system, we obtain
	\begin{equation}\label{eq25}
		\begin{aligned}
			&\sum_{i,\, k}(\mathfrak{s}^{rs}_{ik})^2 = \;\frac{1}{2}\left( 1-\text{cos}\,\theta_{rs} \right)^2 \\
			&\sum_{i}(\mathfrak{s}^{rs}_{i4})^2 - \frac{1}{2}\mathfrak{s}^{rs}_{ik}(e_i^re_k^s+e_i^se_k^r)\; =\; \frac{1}{4}\left( 1-\text{cos}\,\theta_{rs} \right)^2 \, .
		\end{aligned}
	\end{equation}
	Regarding the expression (\ref{eq23b}), we first note that for $ \vec{h}_s $, $ \vec{e}_s $ and $ \vec{k}_s $ as a fixed reference coordinate system with fixed $ \lambda_s $, the (up to now arbitrary) direction of  $ \vec{e}_r $, or $ \vec{h}_r $ respectively, coincides with the line marking the intersection between the planes $ (\vec{e}_s,\vec{h}_s)$ or $ (\vec{e}_r,\vec{h}_r)$ respectively, depending on whether or not $ \lambda_r = \lambda_s $; hence if we set
	\begin{equation}
		\vec{e}_r\bigcdot\vec{e}_s = \text{cos}\,\varphi_{rs} \qquad \text{when}\; \lambda_r=\lambda_s\nonumber
	\end{equation}
	then for a given $ \vec{k}_r $ we have
	\begin{equation}\label{eq25b}
		\begin{aligned}
			(\vec{h}_r\bigcdot\vec{h}_s)^2 - (\vec{e}_r\bigcdot\vec{e}_s)^2 = -\text{cos}^2\varphi_{rs}\text{sin}^2\theta_{rs}\quad\text{when}\; \lambda_r=\lambda_s\\
			(\vec{h}_r\bigcdot\vec{h}_s)^2 - (\vec{e}_r\bigcdot\vec{e}_s)^2 = -\text{sin}^2\varphi_{rs}\text{sin}^2\theta_{rs}\quad\text{when}\; \lambda_r\neq\lambda_s 
		\end{aligned}\tag{25\textquotesingle}
	\end{equation}
	Therefore, we now see how the factors (\ref{eq25}) and (\ref{eq25b}) cancel the singularities in $ I_{\pm\rho}(rs) $ for $ \theta_{rs} = 0 $. 
	
	Using equations (\ref{hamilt-G}), (\ref{eq24}), and (\ref{eq25}), the contribution of $ \mathcal{H}_G $ to the perturbed energy can be written as
	\begin{equation}
		\begin{aligned}
			\overline{\mathcal{H}}_G = \frac{\varepsilon^2\alpha^4}{32\pi^2}&\frac{1}{L^3}\;\sum_{rs}\;\,  \big[ (\text{cos}\,\theta_{rs}-1)  \left\lbrace 2A_rA_sB_rB_s + 2B_rB_sA_rA_s\right.\\
			&\left. - (A_rB_sA_sB_r + A_rB_sB_rA_s + B_rA_sA_rB_s + B_rA_sB_sA_r) \right\rbrace    \\
			&+  \frac{(k_r+k_s)^2}{k_rk_s}\left\lbrace 2A_rA_sB_rB_s+2B_rB_sA_rA_s\right\rbrace  \\
			&+  \frac{(k_r-k_s)^2}{k_rk_s}   (A_rB_sA_sB_r + A_rB_sB_rA_s + B_rA_sA_rB_s + B_rA_sB_sA_r) \big]    \nonumber
		\end{aligned}
	\end{equation}
	or after a separate treatment of the $ r=s $ terms \cite{explain},
	\begin{equation}
		\begin{aligned}
			\overline{\mathcal{H}}_G =  \frac{\varepsilon^2\alpha^4}{16\pi^2}\frac{1}{L^3}\;&\sum_{rs}\;\, [(\text{cos}\,\theta_{rs}+1) (A_rB_r-B_rA_r)(A_sB_s-B_sA_s) \\
			&\qquad\qquad+ \frac{k_r^2+k_s^2}{k_rk_s}(A_rB_r+B_rA_r)(A_sB_s+B_sA_s)] \\
			-\frac{\varepsilon^2\alpha^4}{8\pi^2}\frac{1}{L^3}\;&\sum_{r}\;\,  [(A_rB_r+B_rA_r)^2+(A_rB_r-B_rA_r)^2 - 2A_r^2B_r^2-2B_r^2A_r^2 ]\nonumber
		\end{aligned}
	\end{equation} 
	and finally, according to eq. (\ref{eq16}) \cite{commutators}
	\begin{equation}\label{eq26}
		\begin{aligned}
			\overline{\mathcal{H}}_G &= \frac{\varepsilon^2\alpha^4}{16\pi^2}\frac{1}{L^3}\;\sum_{rs}\;\, (\text{cos}\,\theta_{rs}+1) +\frac{\varepsilon^2\alpha^4}{4\pi^2}\;\sum_{r}\;\, \frac{1}{L^3} \\
			&+ \frac{\varepsilon^2\alpha^4}{16\pi^2}\frac{1}{L^3}\;\sum_{rs}\;\,\frac{k_r^2+k_s^2}{k_rk_s}(2N_r+1)(2N_s+1)\, .
		\end{aligned}
	\end{equation}
	
	Performing a similar calculation, noting that the mean values of $ \text{cos}^2\varphi_{rs} $ and $ \text{sin}^2\varphi_{rs} $ are equal, it follows from equations (\ref{H-int}), (\ref{eq24}), (\ref{eq25}) and (\ref{eq25b}) that the contribution of the interaction term $ \overline{\mathcal{W}} $ vanishes. Hence, eq. (\ref{eq26}) is the final expression for the perturbed energy \cite{interaction}.
	
	Had we considered a \textit{classical} wave packet, we would have obtained a \textit{finite} contribution for the gravitational energy, which can be deduced from eq. (\ref{eq26}) by replacing $ 2N+1 $ with $ 2N $ and deleting the first line  \cite{finite}. From the point of view of quantum mechanics, on the other hand, we end up with an infinite contribution, due to the presence of vibrational modes corresponding to arbitrary short wavelengths, which remain even if we replace $ A_rB_s $ with $ B_sA_r $ in (\ref{em-stress-energy-2}) to eliminate the zero point energy of the radiation \cite{regularization}. This divergent contribution has two parts: one part independent of the number of light quanta and one, also infinite, which is proportional to the number of light quanta \cite{Different-pieces}.
	
	The divergent terms, namely
	\begin{equation}
		\lim_{L\rightarrow\infty}\frac{1}{L^3}\int k^n dk_1dk_2dk_3\quad (n=1,2,3)\nonumber
	\end{equation}
	can be rewritten in a more instructive form, which at the same time shows that the limit $ L = \infty $ is irrelevant. Indeed, if $ u(\vec{k}, \vec{x}) $ denotes a normalized eigenfunction, the normalization factor $ 1/L^3 $ can be written as follows
	\begin{equation}
		\frac{1}{L^3} = u(\vec{k}, \vec{x}) u^*(\vec{k}, \vec{x}) = \int  u(\vec{k}, \vec{x}) u^*(\vec{k}, \vec{x}')\delta(\vec{x}-\vec{x}') dV'\; .\nonumber
	\end{equation}
	Hence
	\begin{equation}
		\frac{1}{L^3}\int k^n dk_1dk_2dk_3=\int\delta(\vec{x}-\vec{x}') dV'\int k^n u(\vec{k}, \vec{x}) u^*(\vec{k}, \vec{x}')dk_1dk_2dk_3\nonumber\; ;
	\end{equation}
	because of the following relationship
	\begin{equation}
		\delta(\vec{x}-\vec{x}') =\int  u(\vec{k}, \vec{x}) u^*(\vec{k}, \vec{x}')dk_1dk_2dk_3\; , \nonumber
	\end{equation}
	and using the notation of Landau and Peierls (\textit{l.c.} p. 189) \cite{Landau},
	\begin{equation}
		\frac{1}{L^3}\int k^n dk_1dk_2dk_3=\int\delta(\vec{x}-\vec{x}') dV'(-\Delta_{\vec{k}})^{n/2}\delta(\vec{x}-\vec{x}') = [(-\Delta_{\vec{k}})^{n/2}\delta(\vec{x}-\vec{x}')]_{\vec{x}=\vec{x}'}\, . \nonumber
	\end{equation}
	It can therefore be said that no finite size can be attributed to a quantum of light. It is hardly necessary to stress the analogy with the case of the electron  \cite{pointlike}.

	\section{First-order interaction processes}\label{section-3} 
	In order to summarize the transition processes triggered by the interaction term $ \mathcal{W} $, we want also to introduce --- along light waves (\ref{em-fields}) --- those gravitational waves, which contribute to our zeroth order approximation.
	
	According to Einstein, there are two types of gravitational waves in vacuum \cite{helicity}. On account of equation (\ref{gauge}), they can be described by the $ \gamma_{11}-\gamma_{22} $ and $ \gamma_{12} $ components when the $ z- $axis is chosen as with the propagation vector. We can then also set $ \gamma_{11}=-\gamma_{22} $ in order to obtain the simplification $ \gamma =0 $. Hence, for an arbitrary wave packet of such waves, we have
	\begin{equation}\label{eq27}
		\gamma = 0\;\quad \text{and}\quad \;\gamma_{i4} =0\; .
	\end{equation} 
	If $ \left\lbrace D^{(r)}_{\alpha\beta} \right\rbrace  $ are the corresponding rotations \cite{rotations} that shift the propagation vector $ \vec{k}_r $ to the $ z- $axis, the remaining $ \gamma_{\mu\nu}\; (\mu,\nu =1,2,3) $ can be written as follows \cite{fourier-gravity}
	\begin{equation} 
		\begin{aligned}
			\gamma_{\mu\nu}=\frac{1}{\pi}\sqrt{\frac{hc}{L^3}}\sum_{\vec{k}_r}\sqrt{\frac{L}{k_r}}\left\lbrace   \frac{1}{2}(D^{(r)}_{\mu 1}D^{(r)}_{\nu 1} - D^{(r)}_{\mu 2}D^{(r)}_{\nu 2})\left( F_{\vec{k}_r}e^{\frac{2\pi i\vec{k}_r\bigcdot\vec{x}}{L}} + F^{\dagger}_{-\vec{k}_r}e^{-\frac{2\pi i\vec{k}_r\bigcdot\vec{x}}{L}}\right)\right. \\
			+\left.\frac{1}{2}(D^{(r)}_{\mu 1}D^{(r)}_{\nu 1} + D^{(r)}_{\mu 2}D^{(r)}_{\nu 2})\left( G_{\vec{k}_r}e^{\frac{2\pi i\vec{k}_r\bigcdot\vec{x}}{L}} + G^{\dagger}_{-\vec{k}_r}e^{-\frac{2\pi i\vec{k}_r\bigcdot\vec{x}}{L}}\right)\right\rbrace  \, .
		\end{aligned}
	\end{equation}
	Writing $ M_{\vec{k}_r,1} $ and $ M_{\vec{k}_r,2} $ to denote the number operator for gravitational quanta of the first and second kind respectively along the direction $ \vec{k}_r $, we obtain
	\begin{eqnarray}
		F_{\vec{k}_r} &=& e^{-\frac{2\pi i}{h}\theta_{\vec{k}_r,1}}\sqrt{ M_{\vec{k}_r,1}},\qquad\qquad  F^{\dagger}_{-\vec{k}_r} = \sqrt{ M_{\vec{k}_r,1}}e^{\frac{2\pi i}{h}\theta_{\vec{k}_r,1}}\nonumber\\
		G_{\vec{k}_r} &=& e^{-\frac{2\pi i}{h}\theta_{\vec{k}_r,2}}\sqrt{ M_{\vec{k}_r,2}},\qquad\qquad  G^{\dagger}_{-\vec{k}_r} = \sqrt{ M_{\vec{k}_r,2}}e^{\frac{2\pi i}{h}\theta_{\vec{k}_r,2}}\; .
	\end{eqnarray}
	Using eq. (\ref{hamilt-G}), the energy of these gravitational waves is \cite{grav-zero-point-energy}
	\begin{equation}\label{eq30}
		\overline{\mathcal{H}}_G = \sum_{\vec{k}_r}\left\lbrace \left( M_{\vec{k}_r,1}+\frac{1}{2}\right) + \left( M_{\vec{k}_r,2}+\frac{1}{2}\right) \right\rbrace h\nu_r\, .
	\end{equation}
	According to eq. (\ref{H-int}), which due to eq. (\ref{eq27}) reduces here to
	\begin{equation}\label{eq31}
		\mathcal{W} = \varepsilon\,\gamma_{\mu\nu}\left( F_{4\mu}F_{4\nu} - \frac{1}{2}S_{\mu\nu} \right) \, ,
	\end{equation}
	to first-order perturbation theory, i.e. probabilities proportional to $ \varepsilon^2 $, only some interactions emerge, i.e. those involving one gravitational quantum and two light quanta \cite{loops}. If, based on well-known arguments, we consider only those processes that conserve the total energy, then the total momentum is also conserved \cite{conservation},
	\begin{equation}
		\left. 
		\begin{aligned}
			k_r &= k_s + k_t \nonumber\\
			\vec{k}_r &= \vec{k}_s + \vec{k}_t \nonumber
		\end{aligned}
		\right\rbrace 
	\end{equation}
	which implies that the three quanta of interest must have the same direction.
	
	If we label the state of a gravitational quantum $ t $ and the states of light quanta $ r $ and $ s $, it follows from eq. (\ref{eq31}) that the transition probability per unit of time equals
	\begin{equation}
		\varepsilon^2c^2h\frac{k_rk_s}{k_t}\frac{1}{L^3}w_{rst}f(N_r,N_s,M_t)\, ;
	\end{equation}
	where $ f(N_r,N_s,M_t) $ represents the usual product of $ N_r $, $ N_s $, $ M_t $; $ N_r+1 $, $ N_s+1 $, $ M_t+1 $; $ N_r+2 $, $ N_s+2 $, depending on the process being considered. Furthermore, if $ \theta_{rt} $ denotes the angle between the polarization direction $ \vec{e}_r $ of the light quantum $ r $ and the specified $ y- $direction in the wave plane of the gravitational quantum, it follows that
	\begin{equation}
		w_{rst} = \frac{1}{4}\text{cos}^2 \left( 2\theta_{rt}\right)  \nonumber
	\end{equation}
	if either the gravitational quantum is of the first kind and the two light quanta involved have the same polarization ($ \lambda_r = \lambda_s $), or the gravitational quantum is of the second kind and the two light quanta have different polarization, otherwise
	\begin{equation}
		w_{rst} = \frac{1}{4}\text{sin}^2 \left( 2\theta_{rt}\right)    \nonumber
	\end{equation}
	for the other two possibilities.

	The transition processes can be described as follows:
	
	\begin{enumerate}
		\item One gravitational quantum vanishes and two (different or equal) light quanta emerge;
		\item Two light quanta vanish and one gravitational quantum emerges;
		\item A light quantum vanishes and another light quantum and a gravitational quantum emerge (frequency decrease of a light quantum);
		\item Light and a gravitational quantum vanish and another light quantum emerges (frequency increase of a light quantum).
	\end{enumerate}
	
	For a cavity, filled initially only with radiation
	(without cool dust!), first-order gravitational effects between the light quanta for the gravitational energy are already sufficient to generate Planck's equilibrium
	(with a speed proportional to $ 1/\varkappa $).
	\newline
	
	I am deeply grateful to Prof. Pauli for his advice and many critical remarks.
	\newline
	
	\textit{Z\"urich}, Physics Institute of the Swiss Federal Institute of Technology, August 14, 1930.

	\section*{Addendum to the revised version}
	Instead of considering, as we did in section \ref{computation}, an initial state for the light quanta with known momentum, one can also compute the expectation value of $ \di \overline{\mathcal{H}_G + \mathcal{W}} $ for the most general wave packet. Let the initial distribution of the state $ r $ be characterized by a (complex) eigenfunction $ \varphi_r(N_r) $ such that:
	$$ \sum_{N_r=0}^{+\infty} \abs{\varphi_r(N_r)}^2 = 1\; . $$ 
	The initial state is defined by specifying arbitrary $ \varphi_r(N_r) $ for all $ r $, with the only condition that the total number of light quanta be a given constant $ N $,
	$$ \sum_{r}\sum_{N_r}\, N_r \abs{\varphi_r(N_r)}^2 = N\; ;$$
	$$\text{where}\qquad \varphi_r(N_r) = 0\qquad \text{for}\qquad  N_r>N\, .$$
	Hence, we have to compute:
	\begin{equation}
		\begin{aligned}
			\text{Expectation}&\;\text{value of}\;\, \overline{\mathcal{H}_G + \mathcal{W}} \\
			&  =  \sum_{N_0,\,N_1\dots}\varphi^*_0(N_0)\varphi^*_1(N_1)\cdots\left( \overline{\mathcal{H}_G + \mathcal{W}}\right) \varphi_0(N_0)\varphi_1(N_1)\cdots\, .
		\end{aligned}
	\end{equation}
	For $ N=0 $ (in the absence of light quanta) we reach, of course, the same result obtained in section \ref{computation} \cite{vacuum}. Let us consider the case of \textit{one} light quantum, i.e.
	\begin{eqnarray}
		\varphi_r(N_r)=0\; for\; N_r>1\nonumber\\
		\abs{\varphi_r(N_r=0)}^2 + \abs{\varphi_r(N_r=1)}^2 = 1\, ,\nonumber\\
		\sum_{r}\,  \abs{\varphi_r(N_r=1)}^2 = 1\, .\nonumber
	\end{eqnarray}
	First, the diagonal elements eq. (26) have to be considered: their contribution is infinite as in section \ref{computation}. Finally, the other elements make only finite contributions, as one can easily conclude.
	
	\vspace{5mm}
	
	\textit{The following absurd statement has been re required by the journal and is not present in the original German paper.}
	
	Data Availability Statement: No Data associated in the manuscript.


\begin{thebibliography}{20}
\bibitem{footnote}
		In the original paper, the footnotes are unnumbered and only the symbols * and ** are used. To avoid confusion caused by the different numbering of the pages, we used Arabic numerals.
\bibitem{retardation}
		Retardation effects occur because, unlike massive electrons, light quanta cannot behave like slow moving particles. Hence, deviation from Newton's action at a distance law cannot be avoided even for weak gravitational effects. They are implied by the Green's functions, represented by $ I_{\pm 6}$ in momentum space, see eq. (\ref{eq20}), needed to integrate the linearized Einstein equations (\ref{eq10}).
\bibitem{Newton-Constant}
		We changed the notation for Newton's constant, which in Rosenfeld's original paper is $ f $.
\bibitem{old-time-coordinate}
		We retained Rosenfeld's old notation $ x^4=ict $. This choice could produce some extra $ i $'s in some results when they are compared with modern QFT textbooks. With this convention, the flat metric tensor is the Euclidean identity matrix $ \delta_{ik} = diag(1,1,1,1) $.
\bibitem{dalembertian}
		We replaced $ \di \sum_r \frac{\partial^2\gamma'_{ik}}{(\partial x^r)^2} $ with $ \square \gamma'_{ik} $, introducing the modern symbol $ \square $ for the d'Alembert operator.
\bibitem{variables}
		Rosenfeld did not introduce the conjugated momenta for electromagnetic and “primed” gravitational potentials. Despite this, his expressions for $ \mathcal{H}_{L} $, $ \mathcal{H}_{G} $ and $ \mathcal{W} $ correspond to the Hamiltonian density obtained using a Legendre transform of eq. (\ref{Lagrangian}) but with the conjugated momenta written in terms of the old variables.
\bibitem{time-derivative}
		We introduced the short notation $ \di \dot{\gamma'} := \frac{\partial\gamma'}{\partial x^4} $. We shall refer to $ \dot{\gamma'} $ as the “time” derivative.
\bibitem{polarization}
		Rosenfeld used again the word \textit{Eigenschwingungen}, which denoted the physical degrees of freedom that can be identified with the two polarization states.
\bibitem{scalar}
		Unlike Rosenfeld, we did not use Gothic letters for $ e $ and $ h $. In our notation $ e^r_i $ is the $ i- $th component of the vector $ \vec{e}_{\vec{k}_r\lambda_r} $, i.e. $e^r_i = (\vec{e}_r)_i $ . Hence, the first line of eq. (\ref{eq18}) can be also be rewritten as $ \di\mathfrak{w} =\frac{1}{2}(\vec{e}_r\bigcdot\vec{e}_s + \vec{h}_r\bigcdot\vec{h}_s)  $. We have corrected an obvious typographical error in the indices of Rosenfeld's eq. (18) (in the third line).
\bibitem{self-interaction}
		With this statement, Rosenfeld emphasized that he did not consider graviton self-interactions. Rosenfeld discarded the contribution of the gravitational pseudotensor $ \mathfrak{t}_{rs} $ in Einstein's equations, which would encode the graviton self-interaction, by considering only the first-order terms. Indeed, the interaction term producing the Feynman diagram representing the three graviton vertex is proportional to $ \varepsilon^2 $ and it does not contribute to the one-loop diagrams of the self-energy.
\bibitem{energy}
		We remember that the symbol $ \overline{\mathcal{H}} $ stands for the Hamiltonian, i.e. the integral of the Hamiltonian density over the spatial three-volume, namely  $\di \overline{\mathcal{H}} = \int \mathcal{H} dV $.
\bibitem{duepunti}
		In Rosenfeld's words, the two expressions on the left side of the colons (:) “are related” to the expression displayed on the right side, which means that they are proportional, as the following equations show.
\bibitem{Different-pieces}
		This observation was emphasized by Wolfgang Pauli in his revision of the manuscript. We shall analyze Pauli's suggestions in a forthcoming paper.
\bibitem{Landau}
		Landau-Peierl's operator $ \sqrt{\Delta} $ acts on a wave function $ \Phi (\vec{x}) $ as follows. Let $ \varphi(\vec{k}) $ be the Fourier transform of the wave function, namely $ \di  \Phi (\vec{x}) = \int \varphi(\vec{k})e^{i\vec{k}\bigcdot\vec{x}}d^3k\; $,
		Landau and Peierls defined
		\begin{equation}
			\sqrt{\Delta}\Phi (\vec{x}) = \int ik \varphi(\vec{k})e^{i\vec{k}\bigcdot\vec{x}}d^3k\; ,
		\end{equation} 
		where $ k = \abs{\vec{k}} $.
\bibitem{fourier-gravity}
		This is the first time that the plane wave decomposition for the gravitational field appears.
\bibitem{grav-zero-point-energy}
		Rosenfeld did not make any comment on eq. (\ref{eq30}). It is worth noting that also the gravitational zero-point energy appears in this expression.
\bibitem{loops}
		This statement confirms that Rosenfeld did not identify the tad-pole process as a first-order contribution, even if its probability amplitude is proportional to $ \varepsilon^2 $.
\end{thebibliography}
\end{document}